Compositional Constraints for *Lucy* Mission Trojan Asteroids via Near-Infrared Spectroscopy

Benjamin N. L. Sharkey
sharkey@lpl.arizona.edu
Lunar and Planetary Laboratory, University of Arizona, 1629 E University Blvd, Tucson, AZ 85719

Vishnu Reddy
Lunar and Planetary Laboratory, University of Arizona, 1629 E University Blvd, Tucson, AZ 85719

Juan A. Sanchez
Planetary Science Institute, 1700 E Fort Lowell, Suite 106, Tucson, AZ, 85719

Matthew R.M. Izawa
Institute for Planetary Materials, Okayama University, 827 Yamada, Misasa, Tottori 682-0193, Japan

Joshua P. Emery
Northern Arizona University, NAU Box 6010, Flagstaff, Arizona 86011-6010, USA,
University of Tennessee, Knoxville, TN, USA

Abstract

We report near-infrared (0.7-2.5 μm) reflectance spectra for each of the six target asteroids of the forthcoming NASA Discovery-class mission, *Lucy*. Five Jupiter Trojans (the binary (617) Patroclus system, (3548) Eurybates, (21900) Orus, (11351) Leucus, and (15094) Polymele) are well-characterized, with measurable spectral differences. We also report a survey-quality spectrum for main belt asteroid (52246) Donaldjohanson. We measured a continuum of spectral slopes including "red" (Orus, Leucus), "less red" (Eurybates, Patroclus-Menoetius) and intermediate (Polymele), indicating a range of compositional end-members or geological histories. We perform radiative transfer modeling of several possible surface compositions. We find that the mild-sloped spectra and low albedo of Patroclus and Eurybates imply similar compositions. Eurybates (~7 wt.% water ice) and Patroclus (~4 wt.% water ice) are consistent with a hydrated surface. Models for Orus and Leucus are consistent with each other and require a significantly more reddening agent (e.g. iron-rich silicates or tholin-like organics). Polymele has a linear spectrum like Patroclus, but a higher albedo more closely aligned with Orus/Leucus, defying simple grouping. Solar system formation models generally predict that the Jovian Trojans accreted in the outer solar system. Our observations and analysis are generally consistent with this expectation, although not uniquely so.

# 1. Introduction (Trojan Surfaces)

Surface composition is a direct method to characterize asteroid groups and link their present state to their origins. As the largest asteroid population co-orbital with a planet, Jovian Trojans represent a major test of solar system dynamical evolution (Emery et al. 2015). From the perspective of the Nice model (Tsiganis et al. 2005, Morbidelli et al. 2005, Gomes et al. 2005), Jovian Trojans are deposited into their current stable orbital arrangement from reservoirs in the outer solar system (Morbidelli et al. 2007, Levison et al. 2011). This can be modeled through dynamical simulations which recreate the orbits of Jovian Trojans as captured from disk populations extending to ~30 AU (Nesvorný et al. 2013).

The bulk density of the binary system (617) Patroclus (1.08 $\pm$ 0.33 g/cm$^3$; Marchis et al. 2006), suggests a significant fraction of ice in its interior (Emery et al. 2015), supporting an origin in the outer solar system. Later density measurements of the Trojan (624) Hektor initially found a similar value of 1.0 $\pm$ 0.3 g/cm$^3$ (Marchis et al. 2014), however, Hektor's density measurements are sensitive to shape-modeling assumptions. A much higher density of 2.43 $\pm$ 0.35 g/cm$^3$ was found through re-analysis that assumed a dumbbell-like contact binary shape instead of Roche ellipsoids (Descamps 2015). The presence of two Trojan systems with significantly different internal structures would complicate hypotheses of a single evolutionary pathway. Furthermore, current surface measurements speak to an absence of volatile and organic compounds, despite their anticipated occurrence in the volatile-rich outer solar system. To understand this discrepancy, near infrared spectroscopy (NIR, 0.7-2.5 μm) has been used to characterize the population properties of Trojan surfaces. Two distinct spectral groups (consistent across both L4 and L5 clouds) align with D- and P-type asteroid taxonomies (Emery et al. 2011), which is also observed in optical colors (Roig et al. 2008, Szabo et al. 2007), and infrared albedo (Grav et al. 2012). Notably, Patroclus is a member of the "less-red" group, while Hektor is a member of the "red" group. The question of whether the spectral groups represent differing formation histories remains unresolved.

Detection of diagnostic spectral absorption features in these groups of Trojans would allow for a detailed mineralogical characterization of their surfaces. The absence of observed features in the 3-μm region by Emery and Brown (2004) implies upper limits for complex organic compounds modeled as laboratory tholins. No spectral absorption bands have been characterized in the range 0.2-4.0 μm (Emery et al. 2011, Dotto et al. 2006, Marsset et al. 2014, De Luise et al. 2010, Wong et al. 2019). Brown (2016) detected spectral curvature between 3.0-3.5 μm in several large less-red Trojans (including Patroclus), which may indicate the presence of ices or organic species and would strengthen the connection between Trojans and a volatile-rich origin. Atmospheric extinction in this wavelength region can limit the definition of the spectral continuum, which is necessary to tie these low signal-to-noise features to a specific hypothesis. Without any compositionally diagnostic features in Trojan spectra, the inverse problem of determining surface compositions from near-infrared spectra (0.7-2.5 μm; "NIR") is inherently degenerate and model dependent.

Emery et al. (2006) observed a spectral emissivity feature in Trojans near 10 μm that most closely matched spectra of fine-grained silicates. Detailed fitting of a model to this feature requires complex scattering scenarios involving opaque grains embedded within a transparent matrix (Emery et al. 2006, Yang et al. 2013), or a fluffy fairy-castle texture (e.g., Vernazza et al. 2012). While the strength of this feature can vary substantially, its shape and position appear similar across the limited sample of four asteroids (Mueller et al. 2010). Regardless of the scattering scenario, anhydrous silicates mixed with elemental carbon has been shown to agree with observations across several wavelength regimes, although no current models claim to be uniquely satisfactory.

NASA's Discovery-class mission, *Lucy* (Levison et al. 2016), will conduct flybys of five Trojans and one main belt asteroid ((52246) Donaldjohanson) that span an order of magnitude in both size (diameter) and rotation period. The physical properties of the targets, including visual albedos, are summarized in *Table 1*. Four of the targets are Trojans in the leading L4 Lagrangrian cloud ((3548) Eurybates, (21900) Orus, (11351) Leucus, and (15094) Polymele) and one is in the trailing L5 cloud ((617) Patroclus). Patroclus is a known binary system (Merline et al. 2002), while Leucus is a candidate binary system based on lightcurve measurements (Buie et al. 2018). With constraints already in place on binarity, size, rotation state, and albedo, we present spectroscopic observations of all six target asteroids.

Our focus is to understand how spectral variations among this group can lay groundwork for quantitative, testable predictions of Trojan surfaces. In this work, we provide spectral slopes and characterize the curvature in the reflectance spectra of the observed spacecraft targets. We model the reflectance spectra to explore the various compositions proposed for Trojan surfaces. Finally, to relate spacecraft and ground-based observations, we formulate hypotheses on the range of surface compositions and textures (grain size) for the *Lucy* targets.

## 2. Observations

Spectral observations were performed using the SpeX instrument (0.7-2.5 µm; Rayner et al. 2003) on the NASA Infrared Telescope Facility (IRTF) in low-resolution prism mode (R ~100). Target trace extraction and wavelength calibrations were performed using the Spextool reduction pipeline (Cushing et al. 2004). Further details of our reduction methodology are given in Sanchez et al. (2013) and Sanchez et al (2015). The six targets were observed from May 2018 through June 2019. Measurements of Eurybates and Leucus were attempted in July 2017 but were unsuccessful. A 3-sigma clip relative to the spectral continuum was performed excluding outliers in the telluric water bands and as detector efficiency drops towards 2.5 µm.

The observational circumstances are given in *Table 2*. All asteroid spectra were acquired with airmass less than 2.0. Telluric corrections were performed during every hour of the observations, using a nearby solar-type standard star. One stable solar analog star, either SAO 93936 or SAO 120107, was observed each night to correct for any non-solar deviations of the standard star. We examined all data collected for each object for internal consistency. Guiding errors, especially on faint targets, can produce discordant spectra which we excluded from our final summed spectra. Additionally, atmospheric diffraction can cause wavelength-dependent slit losses at the blue-end of our spectra if the slit angle is far from the parallactic angle. If slit re-alignments are not conducted at regular intervals, this would manifest as a systematic error in measuring the spectral slope from ~0.7-1.0 microns. No such variation was seen (within uncertainties) in our data.

The slit spectra were collected in relative units, since light throughput through the slit mask is not quantified. Instead, we scale our spectra by extending our measured slope at 0.7-1.3 µm to previously published visual albedos from *NEOWISE* (Grav et al. 2012; Masiero et al. 2011). This choice in scaling is justified by the previously observed linearity of Trojan spectra in this wavelength region and the lack of any absorption features (Fornasier et al. 2004). Reflectance values, scaled to geometric albedo as described above, are given in *Figure 1*. Gray regions in spectral plots indicate areas of telluric absorption bands that were excluded from slope or mixing model fits but indicate the quality of sky background subtraction.

2.1: Results: Spectral Slope

Comparison among measured spectral slopes is more easily visualized in relative reflectance (where spectra are measured relative to the flux recorded at a specified wavelength). The fits to our spectra, with reflectance normalized to unity at 1.635 µm, is given in *Figure 2*. Spectra were divided into three wavelength regimes (0.70-1.32 µm, 1.50-1.77 µm, and 2.10-2.35 µm) to measure spectral slope. If two or more wavelength regimes returned mutually consistent slope values (within $1\sigma$ uncertainties), those regimes were combined to give higher precision. The fit uncertainty depends on the accuracy of using piecewise-linear fits to describe the measured spectra (for example, the slope of Donaldjohanson is formally blue, but depends entirely on the slight downturn from 0.7-1.0 µm). A summary of measured slopes is given in *Table 3*.

Previously, Emery et al. (2011) found that Trojans display a spectral color dichotomy, whose 0.85-2.2 µm colors are plotted with $1\sigma$ uncertainties in *Figure 2* for ease of reference. We find that the spectral slopes of the five Trojans span the entire range of this red/less-red dichotomy, with the slope of Polymele plotting as intermediate between the two groups.

Clear curvature (a change in spectral slope) is seen in the spectra of Orus and Leucus between 1.3-1.5 µm, while slight curvature was observed in Eurybates near 2.2 µm. Orus and Leucus display similar slopes from 0.7-1.3 µm, but have differing slopes beyond 1.5 µm. Leucus displaying a steeper slope *Figure 3* gives the relative reflectance spectrum of Eurybates (normalized at 1.635 µm) compared with atmospheric transmission data available from the Gemini Observatory (Lord, 1992). Our slope for Eurybates of 17.7 $\pm$ 0.4%/µm appears initially inconsistent with Yang and Jewitt (2011), who measured the slope to be 21 $\pm$ 2%/µm. However, the 2σ discrepancy is the result of the curvature measured in our spectrum. Plotting a straight-line fit with the steeper slope (blue line, *Figure 3*) shows an excellent agreement.

*Figure* 4 displays our observed spectrum of Patroclus/Menoetius compared with the color indices reported by Emery et al. (2011) from observations taken in 2000. We find good agreement to their reported NIR spectral slope, but find the spectrum to lack any slope break. Unlike Emery and Brown (2003), we do not detect any features near 2.3 µm, which could have provided a positive identification of hydrated silicates.

The spectrum of Polymele has an intermediate slope that is linear to within our uncertainties. The presence of an Orus-like slope break, however, would remain hidden within this data. Future observations to better constrain the presence of a slope break near 1.3 µm would be beneficial in establishing its NIR reflectance as either more Orus-like (in the case of spectral curvature) vs. Patroclus-like (in the case of no spectral curvature).

While these targets do not constitute a statistically representative sample of Trojans, it is noteworthy that the smallest target (Polymele) and the largest (Patroclus/Menoetius) have similarly linear spectra with mildly red slopes. If *Lucy* is able to determine the collisional histories of these bodies, it would provide a strong test of whether there is an orderly relationship between size and reflectance properties, such as those measured in optical colors by Wong and Brown (2015). We discuss the implications provided by this unique spectrum for Polymele's mineralogy (in tandem with its albedo) in our compositional modeling results (Section 4).

While our spectrum of Donaldjohanson formally allows for a precise measurement of its NIR slope (given in *Table 3*), we note that such a slope is only valid under the assumption that its reflectance spectrum is entirely linear within this region. Without sufficient signal-to-noise to search for absorption bands or slope changes at wavelengths longer than 1.3 μm, we conclude only that this main-belt asteroid is with a neutral NIR slope. Follow-up spectral observations of this target are critical for forming more robust hypotheses about its origins within the main belt, including its relationship with the Erigone family (Zappalà et al. 1995). Mineralogical characterization of Donaldjohanson will require additional observations to constrain due to its low mean SNR (~8).

Ultimately, the presence of two very-red Trojans (Orus, Leucus), two less-red (Patroclus/Menoetius, Eurybates), and one that falls in between (Polymele) suggests measurable diversity among these spacecraft targets. Understanding if the variation in NIR reflectance is suggestive of subtle differences in surface composition or grain size (or both) provides the basis for predictive hypotheses that can be tested by *Lucy*. The investigation of surface scattering models consistent with our observations provides the foundation for our remaining interpretations.

## 3. Constraining Spectral End-Members

*3.1 Methodology:*

Our observations distinguish spectral variations amongst the six *Lucy* mission targets. It is of interest to explore how these variations in reflectance properties correspond to variations in surface composition or grain size. Particularly, we aim to address whether the difference between the reddest objects (Orus and Leucus) and the least red (Patroclus, Eurybates) can be explained by changes in composition, grain size, or both. With Polymele, we have an opportunity to explore whether it is similar to Patroclus (with which it shares a similar slope in relative reflectance) or with Orus and Leucus (whose albedos are within 0.01 of Polymele).

The absence of any spectral features leaves open any mineralogical model that can account for a surface that matches both low albedos and red slopes. Therefore, we do not favor any specific compositional models and do not seek to distinguish between them. Due to the wide separation of observed spectral slopes, small changes (<10%) in slope, which may depend on calibration method, are not resolved within modeling uncertainties via this method for these objects.

To compute the geometric albedos of our targets, we use a radiative transfer model following Hapke formalism (Hapke 2012) and the methodology of Emery et al. (2004). The methodology is summarized below. The single scattering albedo for a grain of a single composition, w, is given as

$$w = S_e + (1 - S_e)\frac{(1 - S_i)\theta}{1 - S_i\theta}$$

Where $S_e$ and $S_i$ are the Fresnel reflection coefficients for external and internal scattering, respectively, and $\theta$ is the internal transmission factor. The coefficients $S_e$ and $S_i$ relate directly to the material indices of refraction $n$ and $k$, and are approximated as (Warell and Davidsson, 2010)

$$S_e = 0.05 + \frac{(n-1)^2 + k^2}{(n+1)^2 + k^2} \; ; \; S_i = 1.014 - \frac{4}{n(n+1)^2}$$

while $\theta$ includes the grain size dependence. For the assumption of spherical grains, $\theta$ takes the form

$$\theta = \frac{r_i + \exp[-\sqrt{\alpha(\alpha + s)} \times 0.9D_{\text{eff}}]}{1 + r_i * \exp[-\sqrt{\alpha(\alpha + s)} \times 0.9D_{\text{eff}}]}$$

$$r_i = \frac{1 - \sqrt{\left(\frac{\alpha}{\alpha + s}\right)}}{1 + \sqrt{\left(\frac{\alpha}{\alpha + s}\right)}}; \; \alpha = \frac{4\pi k}{\lambda}; \; s = \frac{1}{D_{eff}}$$

where the quantity $sD_{eff}$ defines the number of scattering events within a single grain and is set to unity. The variability for this quantity is not addressed in our models but would be necessary to quantify the exact relationship between $D_{eff}$ and the physical size of the regolith.

We adopt a three-component mixing model and assume "salt-and-pepper" mixing (i.e. individual particles of a single species). This defines the single scattering albedo, $w_{mix}$, of the mixture to be

$$w_{mix} = \frac{\sum_i \frac{v_i w_i}{\rho_i D_{eff}}}{\sum_i \frac{v_i}{\rho_i D_{eff}}}$$

where $\rho_i$ is the material density and $v_i$ is the mixing ratio in wt.% of species $i$. To convert the single scattering albedo to the geometric albedo of the mixture, we assume isotropic scattering and neglect the effects of the opposition surge as in Emery & Brown (2004). This sets the phase function at zero degrees, $P(0)$, to unity and the opposition surge amplitude, $B_0$, to zero. The geometric albedo is given as

$$A_p = r_0 \left(0.5 + \frac{r_0}{6}\right) + \frac{w_{mix}}{8}\left((1 + B_0) * P(0) - 1\right)$$

$$r_0 = \frac{1 - \sqrt{(1 - w_{mix})}}{1 + \sqrt{(1 - w_{mix})}}$$

To characterize our models, we employ the *emcee* Markov Chain Monte Carlo (MCMC) package (Foreman-Mackey et al. 2012). Such a fitting procedure is well suited to characterize the distributions of Hapke model parameters (mixing ratios and grain size), as MCMC methods accommodate models with highly correlated, nonlinear parameters without assuming a shape for parameter distributions. MCMC provides a flexible framework for determining fit convergence for a variable number of fit parameters. We found that including models with more than three free mixing components decreased the ability of our fits to converge (requiring longer computation times), without increasing fit quality. We interpret this as the limit to the number of components that can be distinguished by our data. While each component of a given mixing model may have its own grain size (or distribution of grain sizes), we assume that a single effective grain diameter, $D_{eff}$, characterizes the surface as seen by our disk-integrated spectra. This provides a benefit in allowing three dimensional fits that display quantitative trends between component mixing ratios and $D_{eff}$. Allowing two or more grain sizes produced fits of similar quality but did not converge for tested runtimes. All models in this work were performed with twenty walker threads for 15,000 iterations. We discard the first 2000 steps, as these initial steps are influenced by arbitrarily assumed initial model conditions. Convergence of all twenty threads into the same best-fit region was observed to be less than or equal to ~1000 iterations for all models of all target asteroids. Overall fit convergence was determined by running models until they returned smooth, single-peaked posterior parameter distributions. This was typically achieved after 7000-8000 steps.

For a surface with a distribution of grains of different sizes, the parameter $D_{eff}$ typically characterizes the smallest grains, as they tend to dominate scattering (Hapke 2012). However, the relationship between $D_{eff}$ and grain size on a real surface depends on how their shapes deviate from the spherical assumption used to derive the equations as previously given, as well as the number of scattering events within a grain (as previously noted).

We restricted grain sizes to a range of 0.5-1000 μm. These wide limits were chosen to be as unrestrictive as possible in order to explore both upper and lower limits of $D_{eff}$. While the Hapke formalism assumes particle sizes larger than the wavelength of reflected light, none of our best-fit models include grain sizes less than 10 μm, which is well above the valid range. By performing an exhaustive investigation of model parameter space, we are able to directly derive uncertainty ranges (percentiles of likelihood) within our models.

Generally, our modelling was insensitive to variations in visual albedo of ~0.01-0.02. Variations of this scale were accommodated by making small adjustments to grain size or composition (adjustments less than the range of derived uncertainties). By allowing grain size to vary over a wide range, we avoid cases where models are affected by user-imposed limits. For example, we find that artificially increasing the albedo of Eurybates by a factor of 2 does not change the mineral mixing ratios, instead shifting the range of valid grain sizes by a factor of ~3 (with a brighter albedo requiring the presence of smaller grains). This shift represents the difference between the darker objects (Patroclus and Eurybates) and the brighter objects (Orus, Leucus, and Polymele).

*3.2 End Member Selection:*
Constraining the surface composition based on featureless spectra is necessarily degenerate. Our goal is to explore possible composition options and investigate how the *Lucy* targets compare based on the end member assumptions we make. To explore surface variation, we create models from two end-member hypotheses for Trojan surfaces. Drawing upon observations by Emery et al. (2004; 2006) of larger Trojans, we know that there is a silicate emission feature ("plateau") near 10 μm. Therefore, we require that each model include silicate minerals in some form. Without an understanding of the variation of thermal infrared emission amongst our sample (our targets are too faint for mid-IR observations), we cannot distinguish between cases of isotropic scattering from a particulate surface or more sophisticated scattering models.

We constructed our models based on two scenarios. The first is a surface dominated by anhydrous, magnesium-rich silicate glasses mixed with elemental carbon ("Pyroxene Rich Model"). This model is based upon previous models of Trojans and has been demonstrated to reproduce the $0.7 - 2.5$ μm reflectance of both red and less-red Trojans (Emery et al. 2011). These proxy materials do not display silicate absorptions in the NIR and feature spectral slopes which vary according to the ratio of Mg to Fe in the bulk material. We use optical constants of two amorphous silicate glasses, $MgSiO_3$ (hereafter "enstatite") and a 60%/40% mixture of $MgSiO_3/FeSiO_3$ from Dorschner et al. (1995). These materials are referred to as P2 and P6 in Emery et al. 2011, respectively. As a darkening agent, we use optical constants for amorphous carbon from Preibisch et al. (1993). Hereafter, we refer to this model as "pyroxene-rich."

To investigate the potential for anhydrous organic materials on Trojans, we adapt the pyroxene rich model by replacing the iron-bearing pyroxene component with Triton tholins (McDonald et al. 1994). This "pyroxene-tholin" model allows us to investigate the effects of varying the choice of reddening components.

Motivated by the proposed connection between Trojans and outer solar system bodies, the third model (hereafter "hydrated silicates" model) seeks to maximize the presence of hydrated minerals using optical constants of the Tagish Lake meteorite as a proxy (Roush 2003). Both the anhydrous and hydrous models were investigated with and without the inclusion of Triton Tholins as a reddening agent. By investigating a family of solutions that vary based on mineralogical components, we provide an analytic view of how constrained these surfaces are from ground-based spectroscopy. We did not attempt to distinguish which models are better or more likely.

The hydrous and anhydrous models were also chosen for comparison because they represent distinct methods of recreating featureless spectra. In terms of model treatment, the pyroxene-rich case approximates a surface with a combination of bright albedo components (silicates) mixed with a darkening agent (amorphous carbon). A good data fit can be understood as being produced piece by piece. First, the spectral slope is matched by finding the correct combination of silicates. Second, the albedo is fit by introducing a (nearly) spectrally-flat darkening agent (amorphous carbon). Model behavior is similar when one of the pyroxenes is replaced by an organic reddening agent (Triton tholins). In this scenario, grain size is a redundant parameter, as its main effect is in matching albedo, which is already done by the inclusion of the darkening agent. Thus, changes in carbon content are accommodated by adjusting grain size. Cases such as these produce unconstrained posterior grain size distributions.

Using a hydrated silicates model based on the Tagish Lake meteorite, however, shows that grain size acts as a more restrictive parameter. In this case, there is no inclusion of a separate high-albedo component. Instead, slopes are described as varying from a featureless, moderately red-sloped component (Tagish Lake) to a steeply red component (Triton tholins). Furthermore, both of these reddening agents have slopes that depend on grain size. For this scenario, there is less room to create a good spectral slope fit at the expense of a poor albedo fit, as grain size is no longer a superfluous free parameter. For a given spectral slope, there is a unique combination of grain size and reddening agents that best match a spectrum. In this scenario, the role of the darkening agent (amorphous carbon) is as a more minor albedo correction. In this way, the combination of Tagish Lake, tholins, and carbon is an attempt to acknowledge the complexity of meteoritic samples while limiting the number of model parameters.

For Eurybates and Patroclus, a final model of water ice (using optical constants from Warren 1984), Tagish Lake, and amorphous carbon was performed. This model produces spectra with mild red slopes and diagnostic water ice absorptions. We do not detect any such absorptions in our sample (and they largely coincide with telluric water vapor absorptions), but computing these upper limits gives physical context for the signal-to-noise we achieved in our spectra. Attempts to perform three-component fits including water ice for Orus, Leucus, or Polymele found poor results due to too few model degrees of freedom.

## 4. Results: Compositional Modeling

We begin by investigating object-by-object comparisons for each model. *Figure* 5 provides an overview of model results for all five Trojan targets, sorted by decreasing values of NIR spectral slope. *Table 4* provides NIR upper limits (median plus 1$\sigma$) of water ice and organic content in tabular form. One-sigma uncertainties for fit parameters are calculated as the range between the 16$^{th}$ and 84$^{th}$ percentiles in model posterior distributions. The pyroxene-rich model shown in *Figure 5A* displays a compositional distinction between the high albedo objects (Orus, Leucus, Polymele) and low albedo objects (Patroclus, Eurybates). The largest difference can be seen by the ~60 wt.% difference in amorphous carbon content between these two groups. We find that confidence intervals on grain sizes largely overlap, but smaller grain sizes are preferred for objects as their spectral slope increases. We note that upper limits to grain sizes above 800 μm are unconstrained as they are close to our cutoff grain size of 1000 μm.

In the pyroxene-tholin model, we find similar trends as the pyroxene-rich model. Here, the split in amorphous carbon content between high- and low-albedo groups falls to ~40 wt.%. Grain sizes are more loosely constrained and almost entirely overlap across all five objects. The hydrated silicates model finds less discrete compositional differences between high and low albedo groups. Across our sample, hydrated silicate (Tagish Lake) content and amorphous carbon content overlaps between high and low albedo objects. Compositional differences between objects are primarily found in the content of Triton tholins, as they control the spectral slope. Allowable grain sizes are smaller and more restricted across all objects.

In the hydrated silicates model, Polymele lies compositionally intermediate between Orus/Leucus and Patroclus/Eurybates. In the pyroxene-rich and pyroxene-tholins model, however, Polymele's bulk composition groups much more closely with Orus/Leucus than Patroclus/Eurybates. Across all three models, Orus and Leucus return models consistent with each other despite measurable distinctions in their spectral slope. Patroclus and Eurybates are similarly clustered. Therefore, while featureless NIR spectra do not prove given two objects have the same composition, we can place constraints on how dissimilar they are.

A full model error analysis with correctly scaled goodness-of-fit measures would require additional modeling to investigate the nonlinear effects of the uncertainty in visual albedo. Critically, such an analysis would also implicitly assume that our choice of end-members the optimal case for a given compositional class. Since we do not exhaust all possible mineral combinations, we provide reduced $\chi^2$ values which are normalized to unity for the pyroxene-rich model. These values are given in Table 5 and provide context for the relative performance of each model given our assumptions.

The comparisons above ignore correlations between model parameters. *Figures* 6-10 report the results of selected MCMC fits for each Trojan target. Figure 6-8 report two models each for Patroclus, Eurybates, and Orus, respectively, which span the full range of fit quality in our sample. Included in each figure is the measured spectrum (plotted in black in the top panel), with the best-fit model given as a red line. Magenta shading indicates the albedo uncertainties reported from NEOWISE. Gray regions in the top panel indicate areas of telluric absorption bands that were excluded from mixing model fits. Residuals of this model (scaled to uncertainties in the relative reflectance spectrum, not including albedo uncertainty) are given in the central panel. The blue shading in the central panel represents the $\pm 1\sigma$ range. The distributions of allowable mixing ratios and grain sizes (which are the basis of posterior model uncertainties) found via MCMC are given in the bottom panel as two-dimensional histograms of grain size vs. component mixing ratio. The shading of the two-dimensional distributions corresponds to the number of iterations the MCMC algorithm spent within a single bin, as darker regions correspond to "preferred" regions. Contours on the two-dimensional histograms indicate the one- and two-sigma uncertainty ranges. The distribution of allowable grain sizes is also presented as a one-dimensional histogram.

The two anhydrous mixing models for Patroclus are given in *Figure 6A-B*. Both find only loose restrictions on grain size, with the pyroxene-rich modeling favoring grain sizes larger than 500 μm, and the pyroxene-tholin model does not find any constraints. Figure 6B shows that the inclusion of Triton tholins halves the silicate content of best fit models and raises their carbon content. However, incorporating tholins, even at the ~5 wt.% level, adds measurable curvature to model spectra which decreases fit quality from 0.7-1.3 µm. Overall, figure 6A-B shows that anhydrous fits for Patroclus requires both large grain sizes and high amounts of darkening agent—adjusting only one factor is not enough to fit both its relative reflectance spectrum and dark visual albedo. Fit residuals are minimized by the pyroxene-rich model, with residuals of the other models performing worse by a factor of 2.3 (see Table 5).

Fits of the pyroxene-rich and pyroxene-tholin models to Eurybates produces similar quality fits as those to Patroclus. As shown in *Figure 7A-B,* Eurybates also provides a good match to the hydrous mineralogy of Tagish Lake. Using the hydrated silicates model for Eurybates, Figure 7A shows that grain sizes are restricted to ranges less than 500 µm and that the inclusion of Triton tholins are unnecessary to reproduce the spectral slope of Eurybates. Replacing Triton tholins with water ice, shown in Figure 7B, improves fit quality (producing fits as good as the pyroxene-rich model) by better matching the spectral curvature we detect near 2.2 µm. For Eurybates, only the hydrated silicates model produces substantially worse fit residuals than the pyroxene-rich model. Figure 7B shows a preference for grain sizes near 500 µm and finds that the tradeoff between amorphous carbon and Tagish Lake is correlated with grain size (values larger than 500 µm favor Tagish Lake content, while smaller values favor carbon). As much as 7 wt.% water ice can be accommodated for very large grains, with 5 wt.% ice allowed at grain sizes near 500 µm. For Patroclus, the model which includes water ice returns the same grain size preferences with a smaller median value of 2 wt.% water ice with an upper limit of 4 wt.%.

Our modeling also finds this conclusion for mixing ratios to be robust with respect to albedo variations. The range of allowed grain sizes, not composition, depends largely upon albedo. Artificially adjusting the albedo of Eurybates and running the same fitting procedure does not affect allowed mixing ratios but does return different grain sizes (while keeping the grain size distribution's shape the same). Thus, we interpret the compositional mixes we recover for Eurybates to be mainly sensitive to our measured spectral slope.

Orus and Leucus show a strong grain-size dependence on mixing model composition. The best-fit pyroxene-rich model for Orus is shown in Figure 8A. While an overall match to both the albedo and NIR spectral slope is observed, noticeable structure is present in fit residuals, leading to a lower fit quality. As seen in Table 5, for Orus and Leucus the hydrated silicate and pyroxene-tholin models both produce better fit residuals than those returned by the silicate rich model. While the tholin content in *Figures* 8A-B and 9 is constant with respect to grain size (as it is the chief reddening agent in these scenarios), there is a strict tradeoff between amorphous carbon and bulk Tagish content dependent on grain size. Unlike anhydrous mineralogies, which return lower limits for grain sizes, the hydrated silicates model returns upper limits for grain size. In the case that there is a non-silicate reddening agent present on Trojan surfaces, this suggests study of plausible darkening agents may provide a means to distinguish compositions of Orus and Leucus based chiefly on spacecraft measurements of visual albedo and grain size.

A characteristic good fit for Polymele is given in *Figure 10.* Polymele provides a unique spectrum that can be viewed as intermediate between that of Orus/Leucus and Patroclus or Eurybates. While its spectrum is linear, similar to Patroclus, its higher albedo is similar to Orus and Leucus. As seen in *Figure 5*, the pyroxene-rich and pyroxene-tholin models group Polymele more closely with Orus and Leucus in terms of composition, but match Patroclus and Eurybates in terms of grain size. For hydrous silicate models, Polymele groups closer in terms of grain size to Orus and Leucus and sits intermediate between the red and less-red objects in terms of composition. This suggests that despite Polymele's linear spectral slope (as does Patroclus), compositionally it is more likely to lie closer to the redder Orus/Leucus than to Patroclus. A slope break at 1.5 μm similar to Orus/Leucus is not detectable due to the lower SNR achieved for Polymele. The presence of any slope break would constrain this comparison considerably. Table 5 shows little difference between model quality for Polymele due to its lower spectral SNR.

Across the Trojans without a slope break near 1.5 μm (Polymele, Patroclus, and Eurybates), pyroxene-rich and pyroxene-tholin models do not constrain grain size estimates beyond lower limits of ~80-200 μm (depending on the inclusion or exclusion of tholins as a discrete reddening agent). However, these models find the differences in our collected spectra to trace with differences in the relative abundance of reddening and darkening agents. Orus and Leucus require increased content of a reddening agent, independent of grain size. Models including the Tagish Lake meteorite instead find the spectral variation to trace more closely with the abundance of reddening agent and regolith grain size. In either case, the darker surfaces of Patroclus and Eurybates require measurably different scattering models than Orus and Leucus.

## 5. Discussion

The compositions of Jupiter Trojans preclude simple explanation from telescopic observations alone. Surface devolatilization similar to observed activity among Centaurs (e.g. Jewitt 2009) presents one mechanism to reconcile the Trojans' expected derivation from a volatile-rich parent population with the observed lack of ices or hydrated materials (e.g., irradiation weathering as in Wong and Brown 2016, or impact-related erosion as in Yang et al. 2013). Low-level outgassing at some point in the history of the Trojans may explain the non-Maxwellian (i.e., non-collisionally relaxed) distribution of their spin rates (Ryan et al. 2016). Water ice on the surface of any Trojans would not be stable down to meter-scale depths over the lifetime of the solar system (Guilbert-Lepoutre 2014) and would require a replenishment process to exist at present. Therefore, detecting trace water ice on Trojan surfaces would imply either that their surfaces are younger or that Trojans have not always existed in their current orbital configuration. Recent impacts could also excavate volatile-rich material at localized regions of these objects.

Determining whether Trojans contain minerals that form in the presence of water ice or other volatile materials provides another key pathway to explore Trojan formation. Hydrated silicates and organics modeled as pure laboratory samples (which display sharp absorption features in the NIR) have not been detected on Jovian Trojans. However, meteoritic samples, such as Tagish Lake, display assemblages of many hydrated silicates and organic materials without displaying NIR absorptions when measured as a bulk sample (dipping less than 0.01 in absolute reflectance near 3 μm; Izawa et al. 2010). Therefore, the use of Tagish lake as a proxy for the non-volatile surface materials allows for spectral modelling that acknowledges the likely mineralogical diversity on asteroid surfaces, including the possibility of partial masking of absorption features. While spatially unresolved spectra are unable to distinguish between models, our derived distributions for two groups of endmembers can provide upper limits for volatile ices and organic compounds. The absolute measurement of these materials stands as a key science objective towards understanding the history of Jovian Trojans (Levison et al. 2016). *Lucy* will provide a detailed picture of these enigmatic bodies' surfaces for the first time. With ground truth for these bodies on the horizon, there is at present an opportunity to test the predictive capability of telescopic data.

Across the models investigated in this work, the dark albedos of Patroclus and Eurybates combined with their less-red spectral slopes produces compositional models which are consistently unique when compared with the reddest objects (Orus and Leucus) and the intermediate spectrum of Polymele. While increasing grain size is one way to produce a lower albedo for a given composition, our three component models show fits for Patroclus require both larger grain sizes and an increasing amount of darkening material. This suggests that, while grouping objects according to spectral slope is useful from a taxonomic sense, albedo is a necessary constraint as well.

We find that models of Eurybates are similar to Patroclus, consistent with upper limits ~7 wt.% water ice on its surface at larger (>500 μm) grain sizes and decreasing ice abundance allowed with decreasing grain size. On Patroclus, this upper limit is ~4 wt.% for water ice, with the same grain size trends. The spectrum of Eurybates is well described with models including > 90 wt.% unweathered Tagish Lake or amorphous carbon. These findings are consistent with those of Emery and Brown (2003), who found that trace water ice is allowed by the 2.0-3.5 μm spectra of some Trojans.

Although Patroclus returns similar model parameters as Eurybates, models which include Triton tholins or Tagish Lake all produce fit residuals worse than the pyroxene-rich model (as seen in Table 5). Since the primary difference between our spectra of Patroclus and Eurybates is the presence of spectral curvature for Eurybates, we conclude that characterization of nonlinearities in Trojan spectra are vital for distinguishing the surfaces of these low-albedo objects. Eurybates is classified as a C-type in the Tholen taxonomy but is the primary body of a collisional family which contains both P- and C-type Trojans (Fornasier et al. 2007, see particularly their Table 5). Investigating the compositional connections between Eurybates and the P-type Patroclus would provide critical context as to how these taxonomic groups relate. For example, P-type Eurybates family members may simply be unrelated asteroids occupying similar orbital locations, or they may indicate that spectrally heterogeneous asteroids can form from the same original parent body.

Orus and Leucus, while measured to have different spectral slopes from 1.5-2.4 μm, produce models which are statistically indistinguishable. Three component Hapke models do not appear sensitive to their difference in slope. Three-component fits including water ice fail to match either spectrum, implying that upper limits are significantly smaller than what can merely be hidden in the noise of our observations (water ice must be a fourth component and is not required to explain any observed features). For the hydrated-silicates case (where the reddening agent and the darkening agent are distinct from the silicate component), their albedo produces tight grain size constraints of 50-100 μm.

Polymele presents an intriguing, intermediate spectrum that defies clear grouping with the other objects. While it has a similar spectral slope to that of Patroclus, it has the brightest visual albedo at 9%. With the caveat that a slope break similar to Orus and Leucus would not be measurable in our data, Polymele's spectrum is consistent with reddening from C chondrite-like materials alone (with the presence of an additional reddening agent allowed but not required).

We find that the spectral diversity of the *Lucy* targets maps to measurable differences in physical composition across three distinct endmember hypotheses. Orus and Leucus are indistinguishable in our models, as are Patroclus and Eurybates. Depending on the endmembers assumed, Polymele either clusters between Orus/Leucus and Patroclus/Eurybates, or it clusters more strongly with Orus/Leucus. We demonstrate that radiative transfer models of Trojan reflectances over $0.7 - 2.5$ μm can be applied with well-defined uncertainties, providing a method to produce quantitative hypotheses for the compositional variety that *Lucy* will see.

6.  Conclusions

1) We measured a continuum of spectral slopes including "red" (Orus, Leucus), "less red" (Eurybates, Patroclus-Menoetius) and intermediate (Polymele) for the Lucy mission Trojan asteroids we observed. This could indicate a range of compositional end-members or geological histories. Mineralogical models of Polymele typically overlap with the "red" objects, with which it shares a similar visual albedo. This suggests that, while grouping objects according to spectral slope is useful from a taxonomic sense, albedo is an equally necessary and important constraint as well.
2) Solar system formation and evolution models generally predict that the Jovian Trojans accreted in the outer solar system, and were subsequently captured at the time when Jupiter and Saturn passed though a 2:1 mean motion resonance (e.g., Morbidelli et al., 2005). The observations and Hapke model results in the present study are generally supportive of this model (although not uniquely so), with surface compositions consistent of low-temperature aqueous alteration mineral assemblages, carbonaceous matter, and ice providing reasonable fits to the spectral data. Such assemblages are expected to form readily in the ice-rich outer solar system.
3) At present, no extraterrestrial materials in our collections are known to have originated from the Jupiter Trojan asteroids. However, a few carbonaceous chondrite meteorites including Tagish Lake, Wisconsin Range (WIS) 91600, and Meteorite Hills (MET) 00432, have been suggested to originate from D-type asteroids (e.g., Hiroi et al., 2003; Izawa et al., 2015). The Hapke model results including Tagish Lake meteorite suggest that materials similar to the "Tagish Lake-like grouplet" meteorites are analogous to at least some of the Jovian Trojans, as it provides a good match in spectral slope to Eurybates and Patroclus without the inclusion of additional coloring agents. This highlights the value of using meteorites consisting of many phases as a single component in Hapke modeling (e.g., Roush 2003), with Tagish Lake providing a material analogue to the assumed non-ice, silicate-dominant, organic-bearing component of the Jovian Trojan surfaces. A recently reported "comet-like" clast in the CR2 chondrite LaPaz Icefield (LAP) 02342 (Nittler et al., 2019) containing high abundances of organic carbon and sodium may also be a material analogue for some Jovian Trojan surface materials, but there are presently no spectral data for this material with which to evaluate such a potential link further.
4) The results of this study present several hypotheses that can be tested by the *Lucy* mission. The spectrally red surfaces of Orus and Leucus are expected to contain less near-surface hydrated material than the "less red" surfaces of Eurybates and Patroclus, with Polymele being intermediate. The variation in spectral slope (and possibly corresponding hydration) could also relate to apparent surface exposure age. If aqueous alteration products such as saponite, serpentine, magnetite or carbonates are present on the surfaces of the Jovian Trojans, that implies that heat source(s) were available to drive aqueous alteration, which would be a key constraint on both the distribution of short-lived heat producing radionuclides in the outer Solar System, and on the collisional/impact heating of icy bodies.
5) Regardless of endmember assumptions, we find that the spectral diversity of the *Lucy* targets maps to measurable differences in surface composition. The dark and neutral objects, Patroclus and Eurybates, require compositional differences from the remaining objects to explain their spectral slopes. Simply adjusting a grain size parameter does not provide sufficient variation to match any mineralogical model to all five Trojans. This conclusion holds true for hydrated or anhydrous materials, and for variations of ~0.02 in albedo. This point is agnostic to hypotheses of the origins of Trojans and can be seen as a direct prediction of our dataset alone. This framework can be used to constrain surface mineralogy from telescopic data, even when diagnostic compositional information is unavailable.

Acknowledgments

We wish to express our thanks to an anonymous reviewer for providing insightful and substantive comments. This work was supported by NASA Earth and Space Science Fellowship (PI: Sharkey) and Near-Earth Object Observations (NEOO) program grant NNXAL06G (PI: Reddy). We thank the IRTF TAC for awarding time to this project, and to the IRTF TOs and MKSS staff for their support. The authors wish to recognize and acknowledge the significant cultural role and reverence the summit of Mauna Kea has always had within the indigenous Hawaiian community. We are most fortunate to have the opportunity to conduct observations from this mountain.

| Target | Diameter (km) | Rotation Period (hr) | Visual Albedo | Comments |
|---|---|---|---|---|
| Patroclus/ Menoetius | 113/104[1] Equivalent sphere: 140.3 ± 0.9[1] | 103.5 ± 0.3[3] | 0.047 ± 0.003[2] | Binary System[8] |
| Eurybates | 63.9 ± 0.3[2] | 8.73 ± 0.01[4] | 0.052 ± 0.007[2] | |
| Orus | 50.8 ± 0.8[2] | 13.45 ± 0.08[5] | 0.075 ± 0.014[2] | |
| Leucus | 34.2 ± 0.6[2] | 445.732 ± 0.021[6] | 0.079 ± 0.013[2] | Likely Binary System[6] |
| Polymele | 21.1 ± 0.1[2] | 5.8607 ± 0.0005[6] | 0.091 ± 0.017[2] | |
| Donaldjohanson | - | - | 0.067 ± 0.034[7] | Main Belt Asteroid |

| | | | |
|---|---|---|---|
| 1: Buie et al. 2015 | 3: Oey 2012 | 5: Mottola et al. 2011 | 7: Masiero et al. 2011 |
| 2: Grav et al. 2012 | 4: Stephens 2010 | 6: Buie et al. 2018 | 8: Merline et al. 2002 |

*Table 1. Physical properties of the Lucy targets, including visual albedo values adopted from NEOWISE for this study. Patroclus/Menoetius is a known binary system, and Leucus is a binary candidate based on its slow rotation period (Buie et al. 2018). Lucy target asteroids span sizes from tens of kilometers to over one hundred and include both fast and slow rotators. Donaldjohanson, the sole main belt asteroid amongst the targets, remains poorly characterized at present.*

| Target | Patroclus/ Menoetius | Eurybates | Orus | Leucus | Polymele | Donaldjohanson |
|---|---|---|---|---|---|---|
| Dates (UTC) | 2019/06/18 | 2018/07/19 2018/08/21 | 2018/6/17 | 2018/07/19 2018/07/20 2018/08/19 2018/08/21 | 2018/07/19 2018/09/02 | 2018/05/21 |
| V Mag | 16.3 | 16.9 17.0 | 17.0 | 17.9 17.9 17.9 17.9 | 18.7 18.9 | 19.0 |
| Phase Angle | 9.8° | 3.5° 4.0° | 5.1° | 4.2° 4.1° 3.6° 3.9° | 3.9° 6.8° | 15.1° |
| Standard Star(s) | SAO119505 | SAO 190082 SAO 189900 | HD 184892 | SAO 145075 SAO 145075 SAO 144725 SAO 144725 | SAO 212692 SAO 212286 | SAO 157807 |
| # Spectra Used/Collected | 26/28 | 14/14 26/26 | 49/60 | 6/8 16/16 26/26 12/12 | 8/8 12/12 | 16/16 |
| Mean SNR | ~170 | ~41 | ~62 | ~25 | ~12 | ~8 |

*Table 2. Summary of our observations of all Lucy targets, including stars used in correcting raw flux into relative reflectance. All targets were observed at low phase angles (Sun-Target-Observer), with the highest phase observation (Donaldjohanson) at 15.1 degrees. A stable solar analog star (SAO 93936 or SAO 120107) was observed on each night to ensure calibration from flux to reflectance values are independent of night-to-night variations in instrumental response.*

| Target | Spectral Slope(s) (%/μm) |
|---|---|
| Patroclus/ Menoetius | 19.7 ± 0.1 |
| Eurybates | 17.7 ± 0.4 (JH) <br> 37.5 ± 6.3 ( K ) |
| Orus | 50.9 ± 0.3 ( J ) <br> 26.8 ± 0.6 (HK) |
| Leucus | 49.7 ± 0.9 ( J ) <br> 33.9 ± 1.6 (HK) |
| Polymele | 28.7 ± 0.8 |
| Donaldjohanson | −2.4 ± 1.6 |

*Table 3. Derived spectral slopes from 0.7-2.5 μm (normalized to unity at 1.635 μm), given by NIR observational bands in the cases with observed slope breaks. While Patroclus/Menoetius and Polymele both have linear NIR spectra, they differ substantially in both spectral slope and visual albedo. Orus and Leucus have similar slopes at J band (1.2 μm), with Orus breaking towards a flatter slope than Leucus at wavelengths greater than 1.5 μm (H and K bands). For Eurybates, we find a ~3σ slope break at 2.2 μm.*

| Target | Water Ice (Upper Limit) | Tholin Organics (Upper Limit) |
|---|---|---|
| Patroclus/Menoetius | 4 wt.% | 5 wt.% |
| Eurybates | 7 wt.% | 7 wt.% |
| Orus | - | 28 wt.% |
| Leucus | - | 28 wt.% |
| Polymele | - | 21 wt.% |

*Table 4. Derived upper limits for water ice and organic content of from three-component Hapke models. Water ice upper limits are calculated assuming endmembers of Tagish Lake, water ice, and amorphous carbon. Tholin upper limits are calculated assuming endmembers of enstatite, Triton tholins, and amorphous carbon.*

| Endmember Model | Patroclus | Eurybates | Orus | Leucus | Polymele |
|---|---|---|---|---|---|
| Pyroxene-Rich | 1.0 | 1.0 | 1.0 | 1.0 | 1.0 |
| Pyroxene-Tholin | 2.3 | 1.1 | 0.3 | 0.5 | 0.8 |
| Hydrated Silicates | 2.3 | 2.5 | 0.6 | 0.8 | 0.9 |
| Water ice Tagish Lake Amorphous Carbon | 2.3 | 1.2 | - | - | - |

*Table 5. Reduced $\chi^2$ values for all models considered in this work, normalized relative to the values for the pyroxene-rich model. Models with the lowest residuals for a given object are presented in bold.*

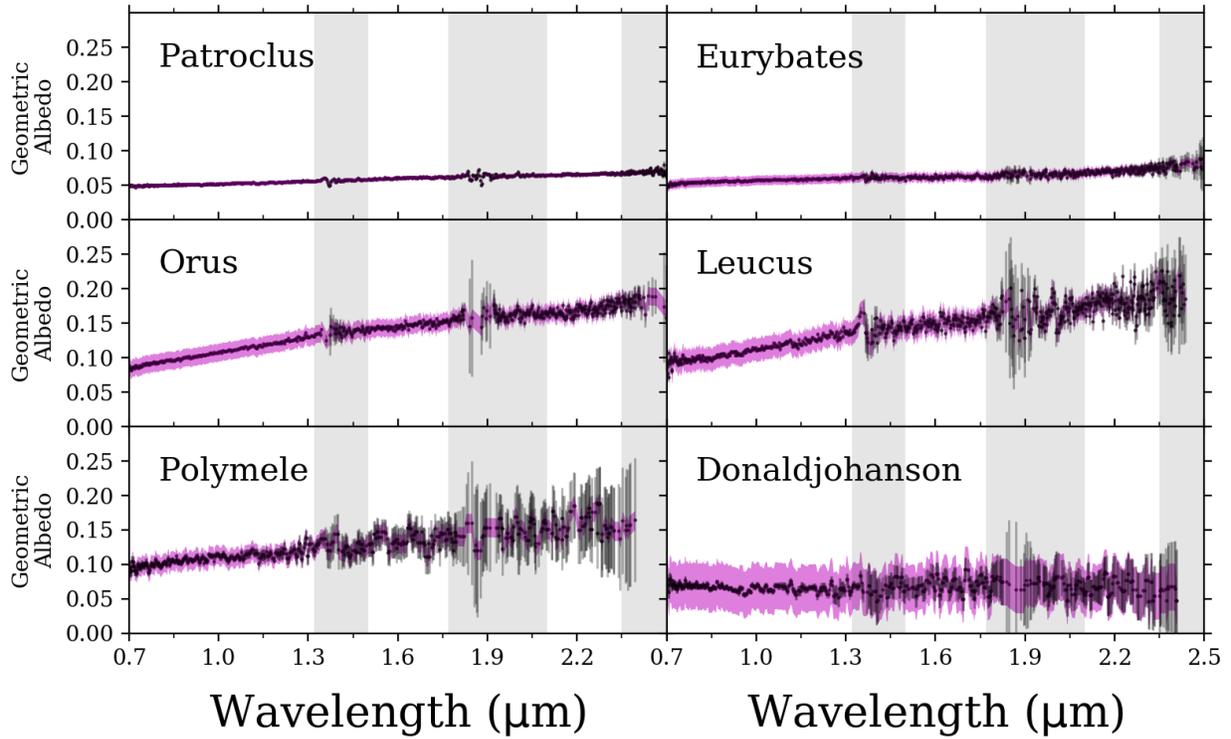

*Figure 1. Near-infrared (0.75-2.5 μm) spectra of all six asteroid targets of the Lucy mission obtained using the SpeX instrument on the NASA IRTF on Mauna Kea, Hawai'i. Conversion from relative reflectance to geometric albedo was performed by scaling to visual albedos as reported by NEOWISE (Mainzer et al. 2016, Grav et al. 2012, Masiero et al. 2011). Relative uncertainties in our spectra are given by the black error bars, the uncertainties of the NEOWISE measurements (corresponding to a vertical translation of our data) are given by the magenta shaded regions. The gray bars indicate the location and extent of the atmospheric water bands.*

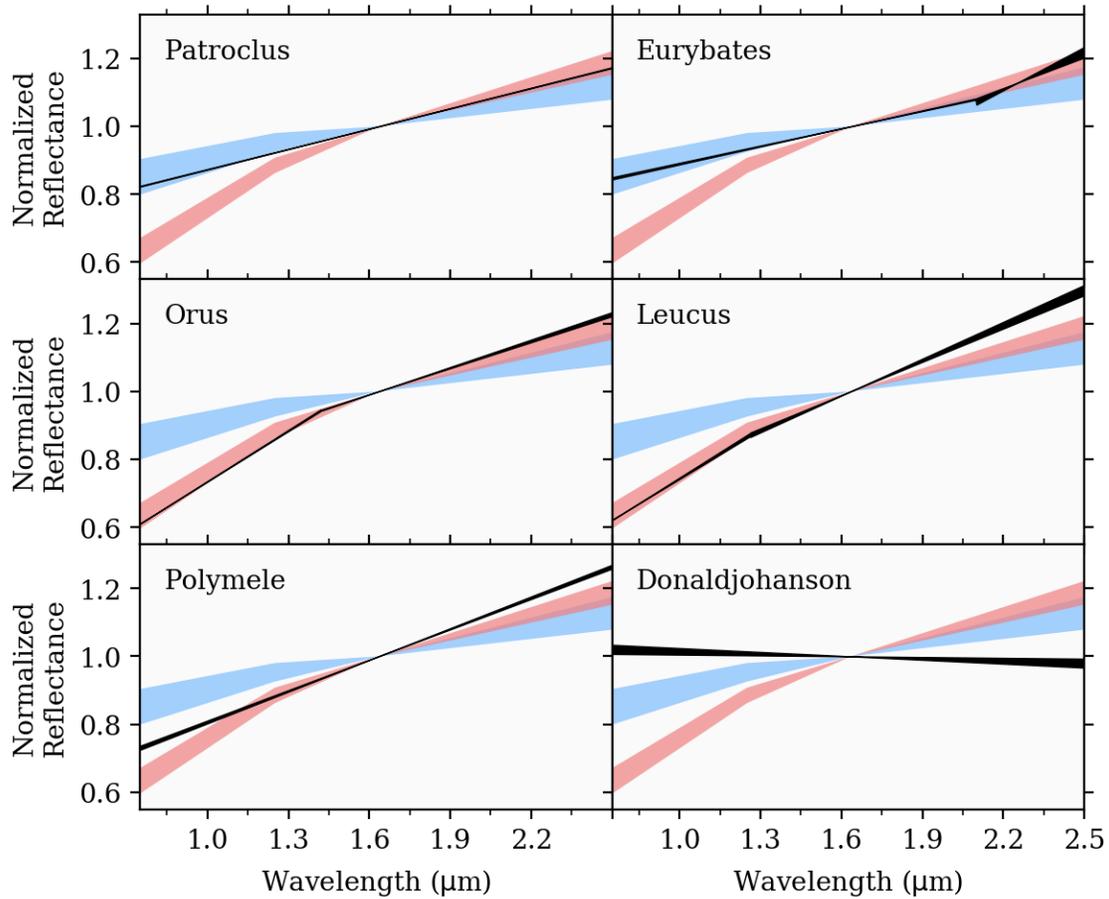

*Figure 2. Comparison of slopes (thicknesses correspond to 1σ uncertainty range) of the all Lucy target spectra in our study. For each panel, the slope of the labeled target is plotted in dark black. Distributions (1σ) of near-infrared colors corresponding to the "red"/"less red" Trojan color dichotomy from Table 2 of Emery et al. (2011) are plotted as light red and blue shaded regions. Reflectance values are normalized to unity at 1.635 μm. Our observations show a continuum of NIR spectral slopes from "red" to "less red" for the Trojan targets. Main belt asteroid (52246) Donaldjohanson displays a neutral slope.*

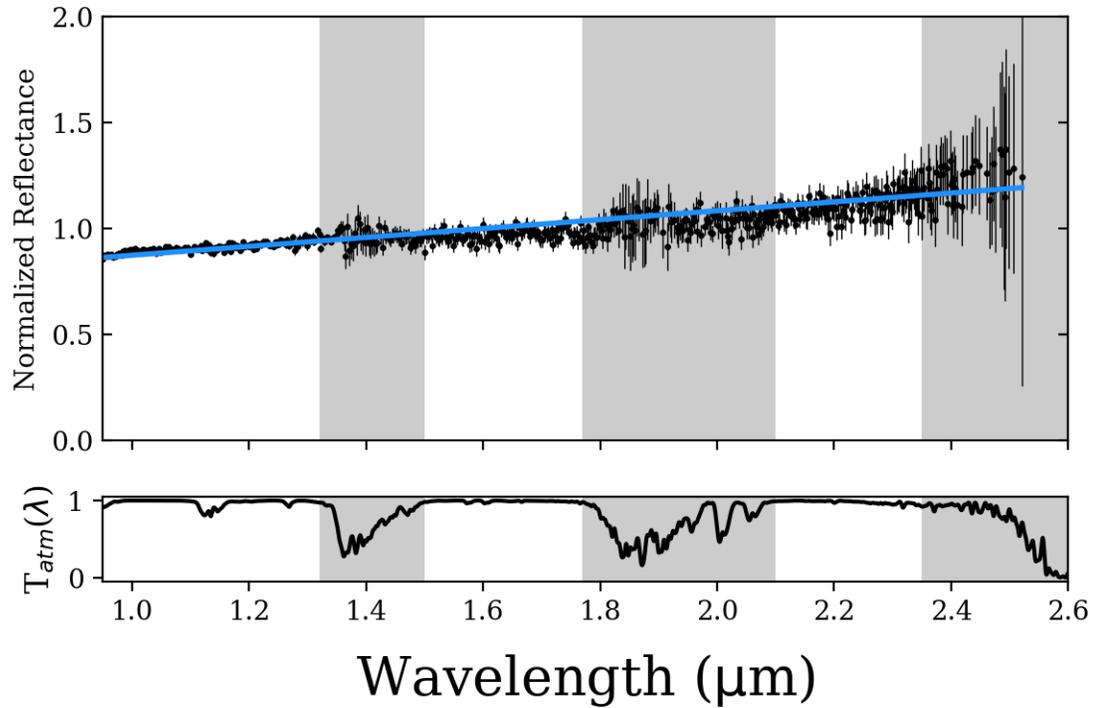

*Figure 3. Reflection spectrum collected for asteroid Eurybates (top), normalized to unity at 1.635 µm. The bottom panel shows atmospheric transmission (Lord, 1992) smoothed to the spectral resolution of our observations. The blue line is the nominal slope measured by Yang and Jewitt (2011), showing good agreement with our observation. Gray boxes indicate regions excluded from mixing model fits and spectral slope fits due to lower signal to noise, although their inclusion in fits does not affect results.*

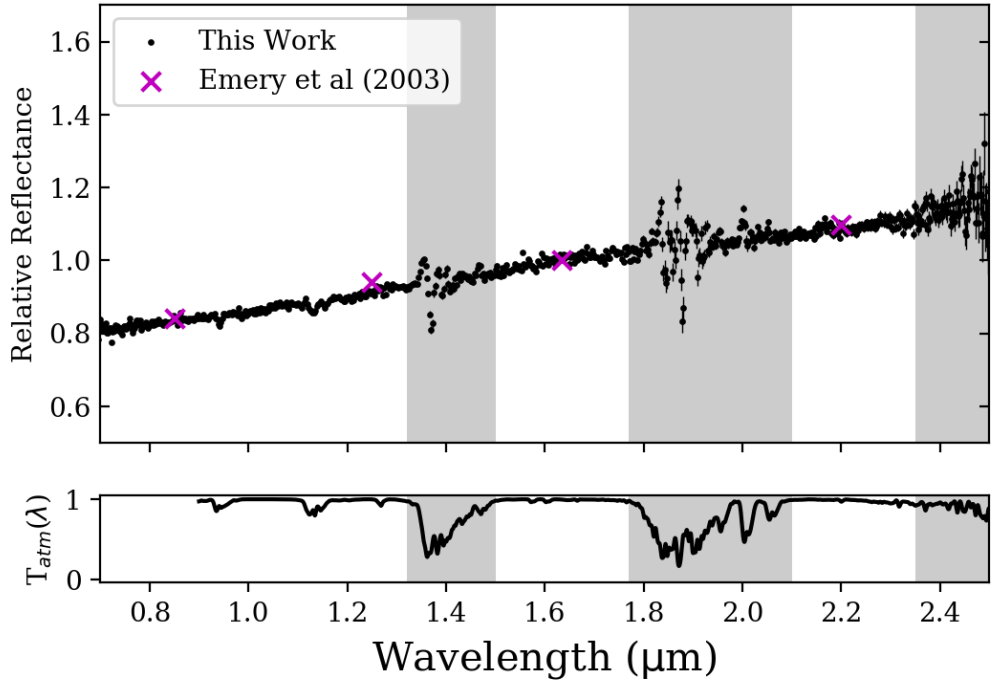

*Figure 4. Comparison of our spectrum of Patroclus/Menoetius with colors derived from Emery et al. (2003). Reflectance values are normalized to unity at 1.635 μm. We find good agreement in overall slope measurements between the 2019 observations and results from Emery et al. (2003). We do not detect any slope break and instead find the NIR spectrum of Patroclus/Menoetius to be strongly linear. Reported uncertainties from Emery et al. (2003) plotted but not visible at this scale.*

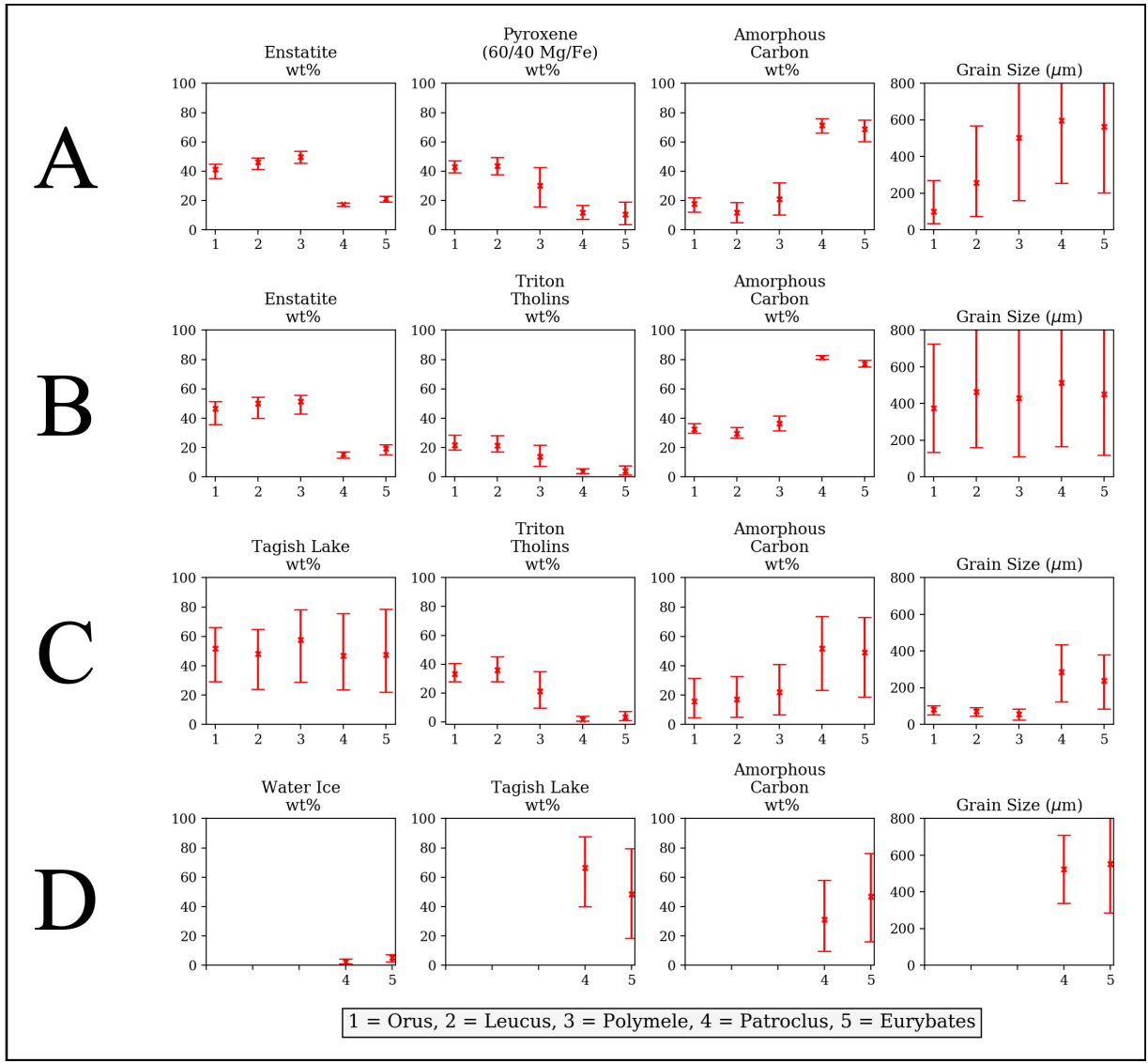

*Figure 5. Summaries of Hapke mixing model results from (A) a pyroxene-rich mixture with amorphous carbon as the darkening agent; (B) enstatite with organics in the form of Triton tholins as a reddening agent and amorphous carbon as a darkening agent; (C) Tagish Lake with Triton Tholins and amorphous carbon. (D) Water ice, Tagish Lake, and amorphous carbon. Mixing ratio uncertainties (1-sigma percentile ranges) are marginalized over all grain sizes. The models in (A) and (B) which include orthopyroxenes place lower limits for grain sizes and find compositional differences between the reddest objects and the least red. The models in (C) which include Tagish Lake find that grain size effects can largely control the differences between Patroclus and Eurybates and the higher albedo objects Orus, Leucus, and Polymele. The two models in (D) are used to return upper limits for water ice content on Patroclus and Eurybates, which are given in Table 4. Models for Donaldjohanson were performed but are not presented, as the the low SNR (mean ~8) achieved by our observations do not constrain our fits.*

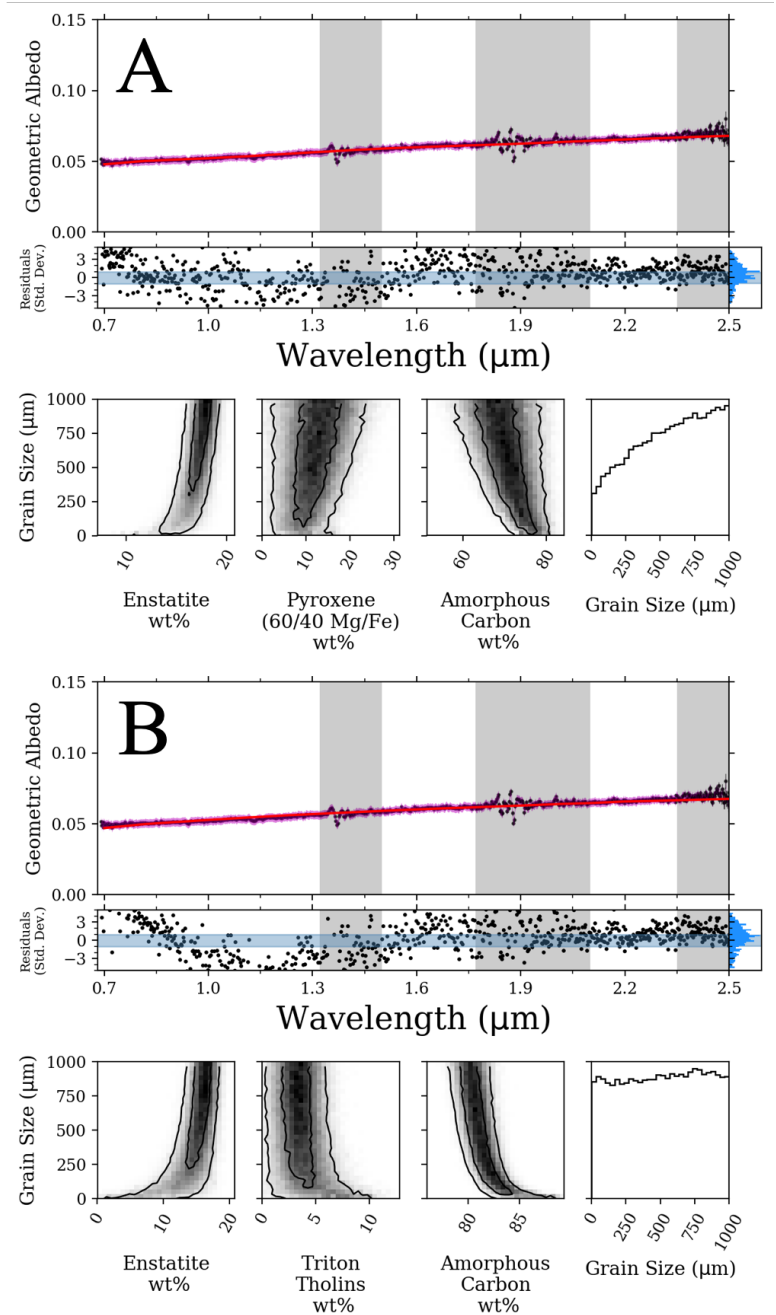

*Figure 6. Comparison between two Hapke mixing models for Trojan binary asteroid Patroclus/Menoetius. Magenta shading indicates the albedo uncertainties reported from NEOWISE. (A) Best fit and model parameters for pyroxene-rich fits to Patroclus. (B) Best fit and model parameters for pyroxene-tholin fits to Patroclus. Both models find a surface dominated by dark, spectrally neutral carbon material. The inclusion of <10 wt.% organics in the form of Triton tholins adds slight curvature to the spectrum which is inconsistent with our measured spectrum. In the pyroxene-rich case, grain sizes that allow higher abundances of carbon are preferred. When Triton tholins are included, grain size controls the trade-off between enstatite and amorphous carbon content.*

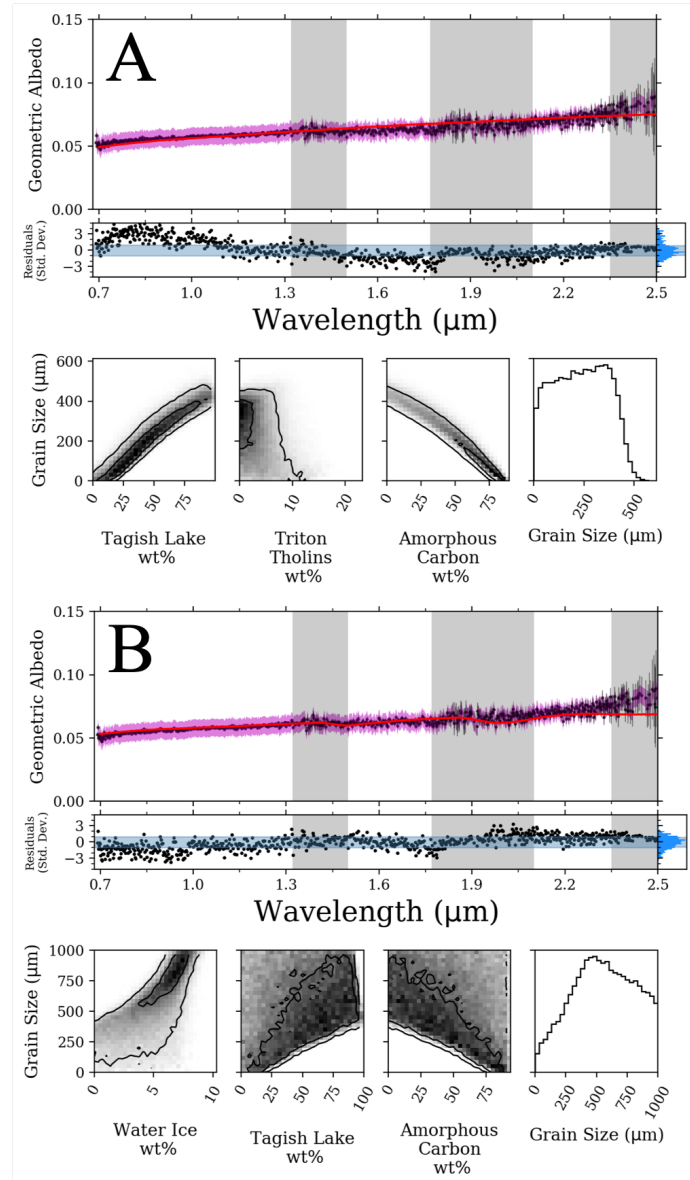

*Figure 7. (A) Best fit and model parameters for hydrated silicate fits to Eurybates. (B) Best fit and model parameters for fits including water ice, used to derive upper limits for allowable water ice content. Top Panels: Comparison of our spectrum of the Trojan asteroid Eurybates (black points) with hydrated models including Tagish Lake (red line), with albedo uncertainties (magenta shading). Central Panels: Residuals of the model fit scaled to data 1-sigma uncertainties, with the ± 1-sigma levels indicated by blue shading, with the distribution of residuals indicated by the horizontal histogram. Bottom Panel: Two-dimensional histograms of mixing ratios showing the uncertainty distributions of MCMC fits, with contours indicating the one- and two-sigma uncertainty ranges. The neutral slope of Eurybates closely matches that of Tagish and amorphous carbon for a wide range of large grain sizes beginning near 375 μm. We find that as much as ~7 wt.% water ice is consistent with our data across a variety of models, with increasing allowances at higher grain sizes which suppresses absorption bands. The inclusion of water ice modestly improves fits near 2.2 μm vs. two component models.*

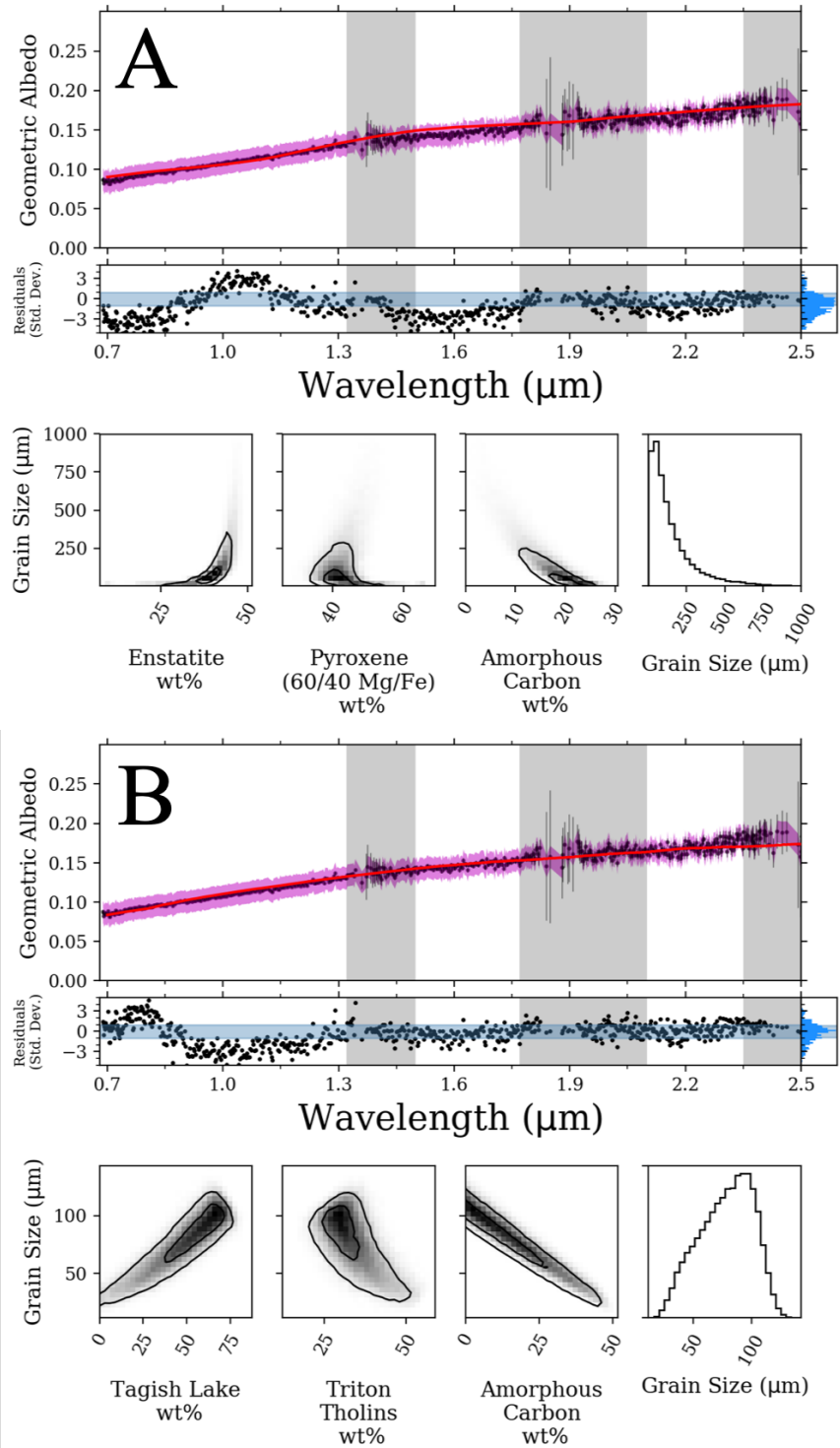

*Figure 8. Comparisons of near-IR spectrum of Trojan asteroid Orus with (A) the pyroxene-rich case and (B) the hydrated silicate case. Magenta shading indicates the albedo uncertainties reported from NEOWISE. Grain sizes are constrained to be smaller than approximately 100 μm. Changes in the grain size directly control the contribution from Tagish vs. amorphous carbon (a darkening agent) in these mixing models.*

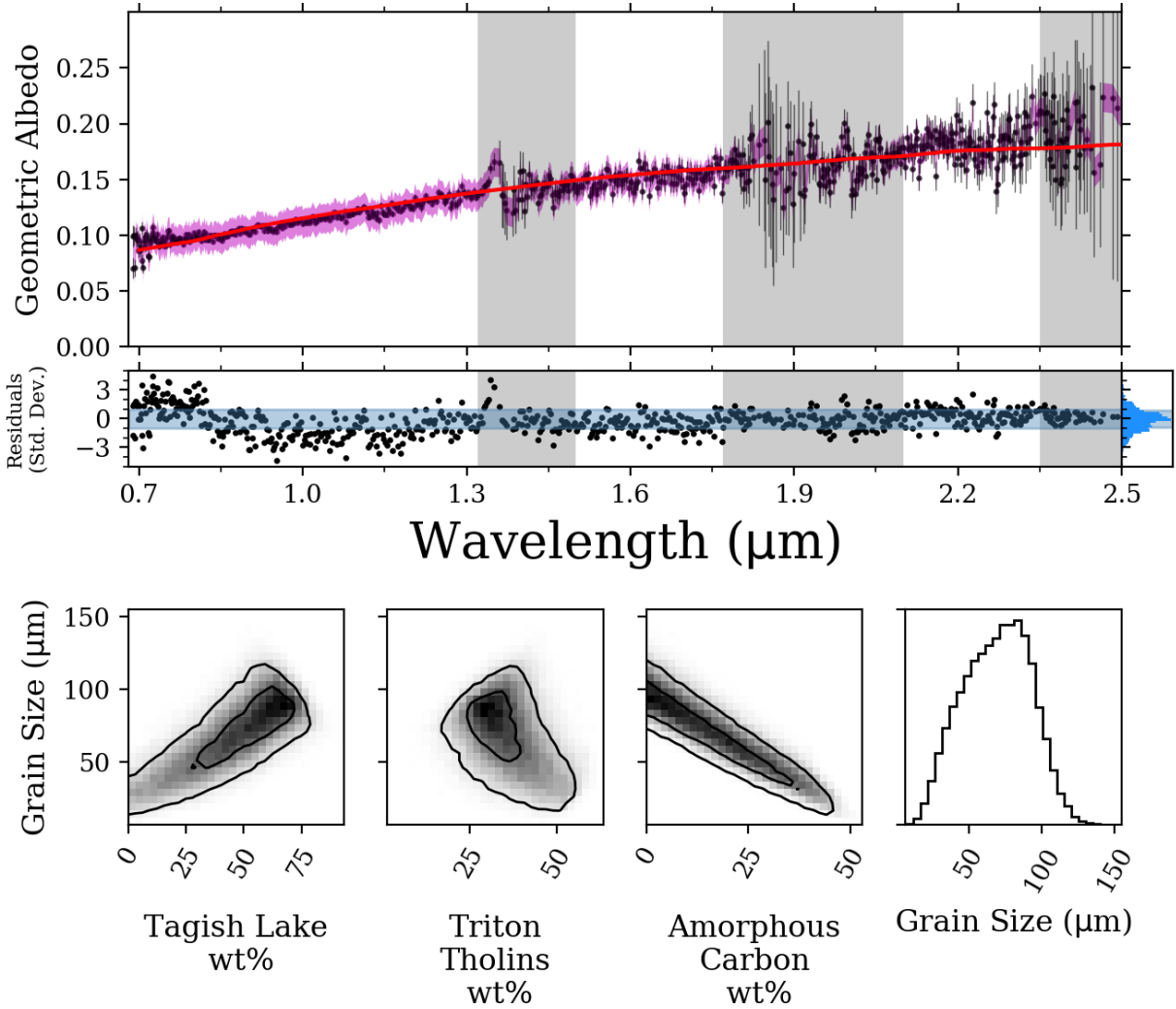

*Figure 9. Comparison of spectra Trojan asteroid Leucus with best fits for the hydrated silicates model. Magenta shading indicates the albedo uncertainties reported from NEOWISE. Grain sizes are tightly constrained to approximately 50-100 μm. Mixing ratios are highly similar to those for Orus, with a preference towards slightly smaller grain sizes accommodating the differences in slope between these two objects.*

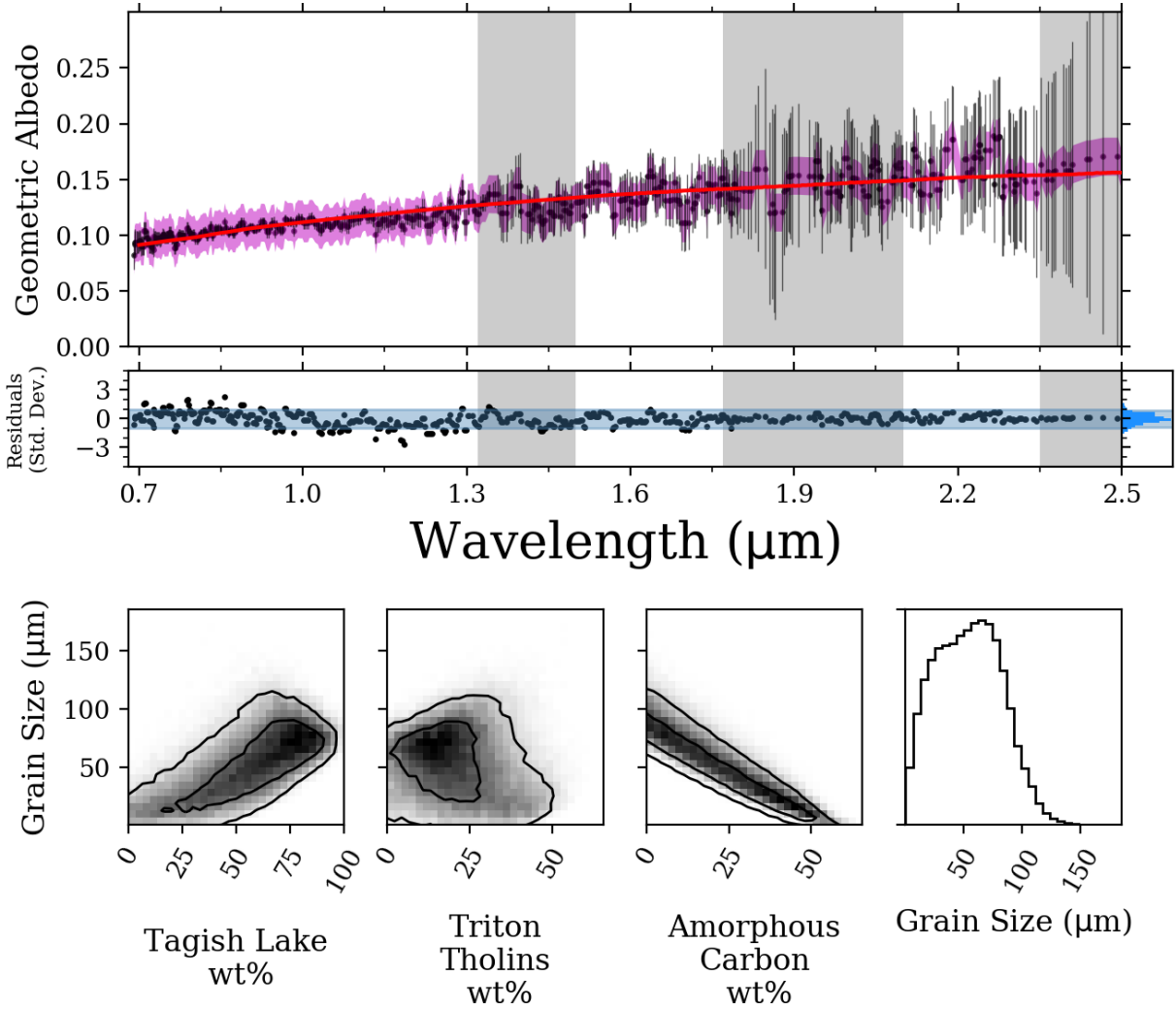

*Figure 10. Comparison of spectra of Polymele, the smallest Lucy Trojan target with with best fits for the hydrated silicates model. Magenta shading indicates the albedo uncertainties reported from NEOWISE. Grain sizes are constrained similarly to Leucus, with less organics in the form of Triton tholins allowed, despite the higher uncertainties in the spectrum of this faint object.*